# Evaluation of Semantic Metadata Pair Modelling Using Data Clustering

Hiba Khalid[1,2], Esteban Zimanyi[2], Robert Wrembel[1]

[1] Poznan University of Technology , Poznan, Poland
[2] University Libre de Bruxelles, Brussels, Belgium
Hiba.khalid@ulb.ac.be, esteban.zimanyi @ulb.ac.be, Robert.wrembel@cs.put.poznan.pl

**Abstract.** One of the most expensive data related tasks is cleaning noisy or messy data. Each dataset comes with specific and general pieces of information. This information could be both relevant or irrelevant to other datasets. To comprehend the connection between two disintegrated datasets; a middleware is required. Metadata presents such medium for connection, elaboration, examination and comprehension of relativity between two datasets. Metadata $\boldsymbol{MD_i}$can be enriched to calculate the existence of connection $\boldsymbol{C}_{(\boldsymbol{di},\boldsymbol{dk})}$ between different disintegrated datasets. In order to do so, the very first task is to attain a generic metadata representation for domains. This representation narrows down the metadata search space $\mathrm{S_i}$. The metadata search space consists of attributes, tags, semantic content, annotations etc. to perform classification. The existing technologies limit the metadata bandwidth i.e. the operation set for matching purposes is restricted or limited. This research focuses on generating a mapper function called 'cognate' $\mathrm{CO_r}$ that can find mathematical relevance based on pairs of attributes between disintegrated datasets. Each pair is designed from one of the datasets under consideration using the existing metadata and available meta-tags. After pairs $\boldsymbol{P}(\boldsymbol{vd_i}), (\boldsymbol{vd_k})$ have been generated, samples are constructed using different combination of pairs. The similarity and relevance between two or more pairs is attained by using data clustering technique to generate large groups from smaller groups based on similarity index. The search space $\boldsymbol{S_i}$ Is divided using a domain divider function and smaller search spaces are created using relativity and tagging as main concept. For this research the initial datasets have been limited to textual information. Once all disjoint meta-collection(X1,…,Xn) have been generated the approximation algorithm calculates the centers of each meta-set. These centers serve the purpose of meta-pointers i.e. a collection of meta-domain representations. Each pointer can then join a cluster based on the content i.e. meta-content. All centers are then pooled through a domain channel and linear pointers are sent across meta-pointers. These linear pointers can then bind or leave a reference tag to the visited meta-node $\boldsymbol{Vmn_j}$. . Each visit is them recorded as a graph communication. The model designed facilitates the use of existing metadata to derive meta-operations and provide evidence of connection or disintegration between domains. It also facilitates the process of possible synonyms across cross-functional domains such as sports and food datasets still might share common attributes, people and details. This can be examined using meta-pointers and graph pools.

## 1 Introduction

The world of datasets is complex and required adequate comprehension for solving problems like data integration. Data integration is a very high in demand and crucial process required by almost all business to perform more advanced business intelligence tasks. Even when the domain context is not business, the need to understand and combine data is very important. The task however is not simplistic in nature. It requires a lot of computation power, machine intelligence and manpower to accomplish this task i.e. data integration or data matching. Numerous scientific studies have been conducted to attain a generalized version for performing integration. However, not much success is achieved that is both generic in concept and computationally stable and cost effective as well. To explore the idea of combining or integrating datasets another important character in the filed observation is metadata. Metadata can provide insight into data, its components and its nature provided it is available, in coherence with quality or can be generated using data profiling for instance. All this available metadata can be used in its crude form or can also be enriched to par take in more complicated data perception and resolution tasks such as data integration and entity resolutions.

In context of highly disintegrated and complex big data, the metadata on its own can provide effective middleware capacities for establishing a knowledge bridge between the disintegrated or heterogeneous datasets. Moreover data provenance can provide high quality legacy details that can identify further relevant metadata or previous derived data sources. Thus, in effect minimizing the overall cost of query processing and information retrieval over traditional databases, semantic databases and graph databases.

## 2 Methodology & Experiment Design

Metadata depends upon the domain or origin of the data. A general purpose systems can be designed using a generalized metadata functionality for example the use of customer process as a metadata class. Metadata $MD_i$can be enriched to calculate the existence of connection $\boldsymbol{C}_((\boldsymbol{di},\boldsymbol{dk}))$between different disintegrated datasets. Usefulness of this approach can be explained in broad terms as follows:

- Use of existing metadata to drive semantic search
- To provide a connection between disintegrated datasets using modelling strategy
- Reference tagging for generating meta-structures from domain pairs

'Cognate' can find mathematical relevance based on pairs of attributes between disintegrated datasets. Each pair is designed from one of the datasets under consideration using the existing metadata and available tags. $\boldsymbol{P}(\boldsymbol{vd_i}),(\boldsymbol{vd_k})$ Pair generation leads to sampling. The similarity and relevance calculated by using data stream clustering technique with similarity indexing. The experiment design was executed on a pool of 18 datasets from the domain of Transport. The column and text similarities have been conducted to obtain the results and ideology on the participation role of metadata as an integral entity in improved data integration. The datasets used are from UK GOV, Kaggle and Eurostat on topics of (transport, accidents, labor surveys and road awareness).

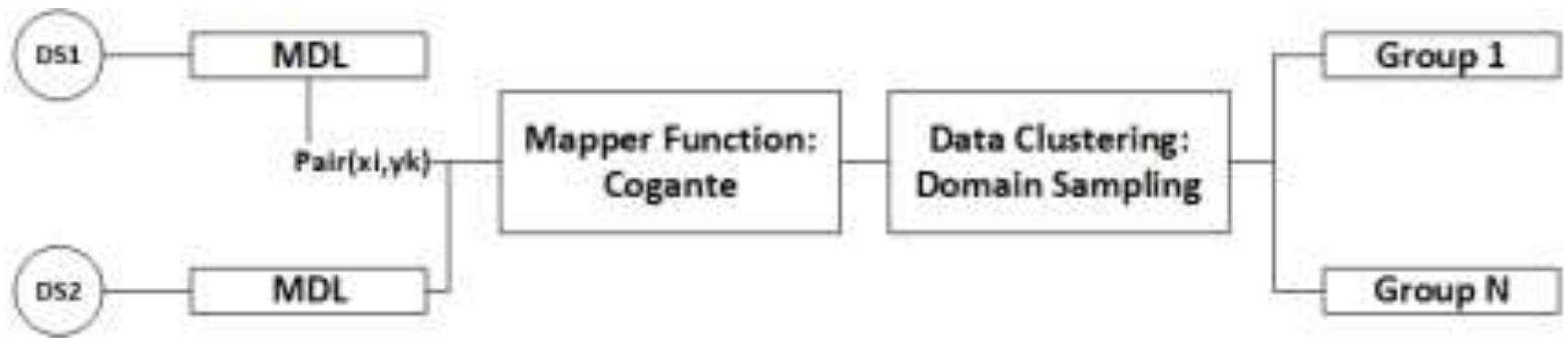

**Fig. 1.** Mapper Function Cognate: This function receives input from different metadata libraries of various data sources. Thus from one to 'n' data sources can provide metadata as input to the function. The cognate function provides mappings for different types of metadata such as administrative, structural, descriptive etc. For instance the mapper can take column names as text input and run a similarity index sampling to provide relevance decisions. It can also take text blocks such as descriptions etc. to create data samples using clustering techniques and define column groups, text groups or domain groups.

### 2.1 Experiment design

The experiment has been designed to accomplish the crucial task of creating pairs using metadata modelling, generating clusters that correspond to a similar domain using a similarity matrix for indication and understanding. The following steps elaborate the most significant procedures required for accomplishing the said scientific task under discussion.

- Create a "Mapper Relevance" class $M_{ci}^{r}$ to locate and create a bulking Id $Bid_{cast}^{domain}$ for similar and relevant datasets.
- The use of "Regression Similarity Technique" for the classifying input data i.e. metadata in bulk.
- The application of "Divider Function $df_{j,k}$" to provide and create a smaller search space i.e. (Si, ……, Sn).
- The calculation of meta-centers $MD_{cd}^{c(i,k)}$ i.e. the focal or the central point in a metadata cluster using annealing approximation.
- The next step is the calculation of meta-pointer pooling using domain classifiers as decider functions.
- Creating and maintaining a logging system called "Linear Point Logging" for maintaining history of nodes i.e. metadata nodes visited for a search cycle $S_{cycle(n-1)}\langle D_{space,i} | D_{space,l}\rangle$

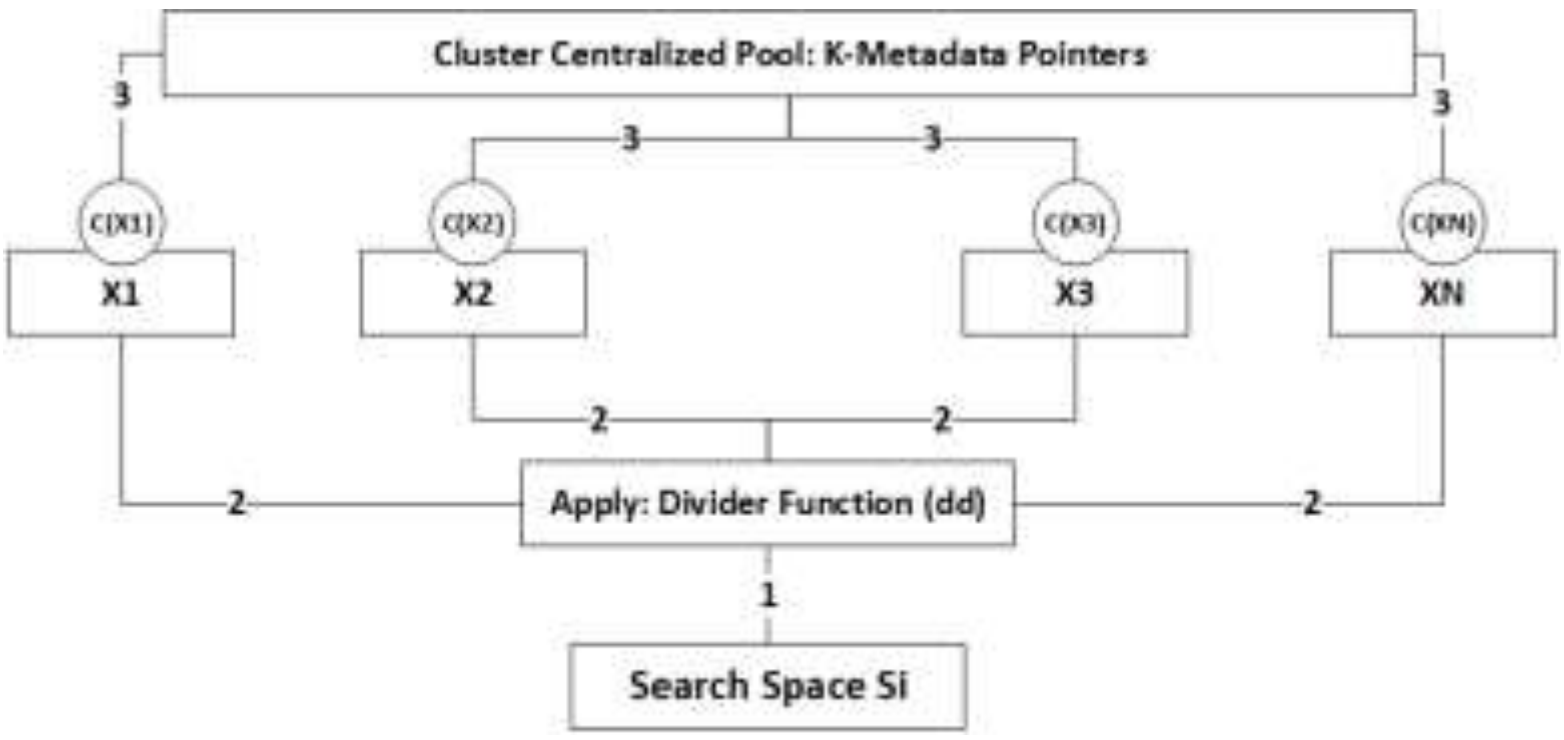

**Fig. 2.** Overall Data Approximation Clustering for Metadata Pointer Pairs.

The resolution strategy presented provides promising results in context of metadata as a catalyst and a contributing entity for data integration cases. The cognate function provides a representation or triple sets for two column comparisons such as (c1,cc1,10). This represents that column 1 C1 od DS1 is 10% similar to column 2 C2 of DS2.

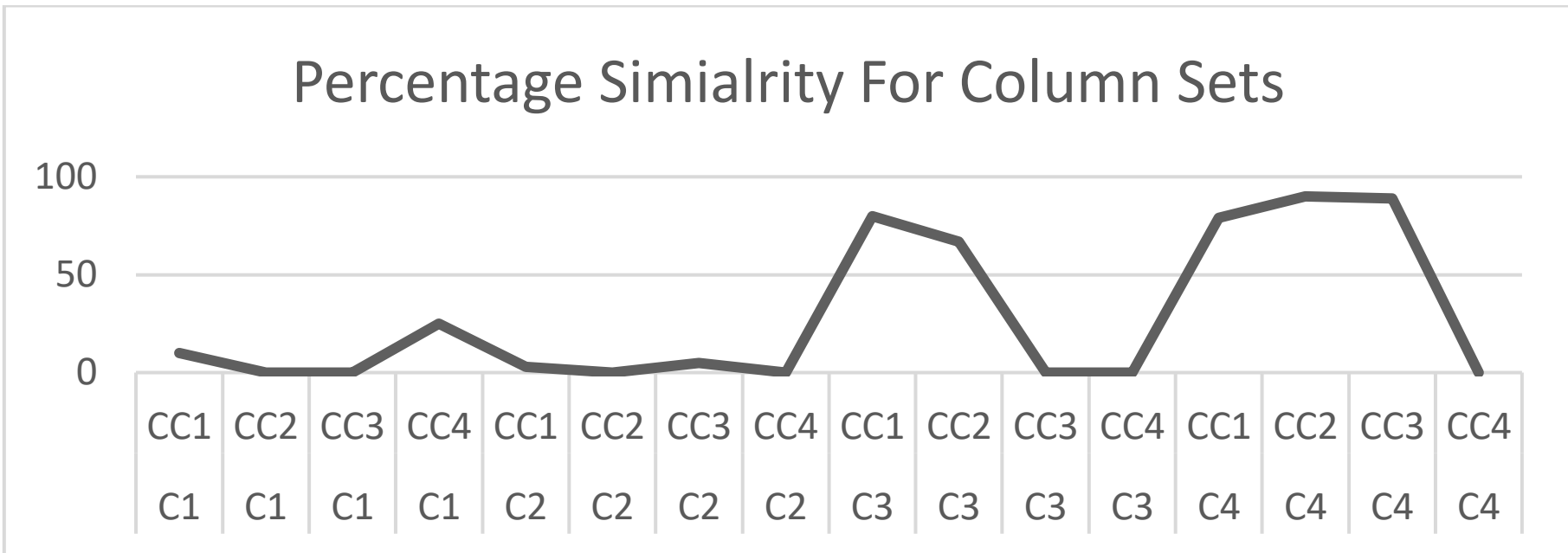

**Fig. 3.** Similarity percentages between columns of two data sources for domain transport & accidents.